\documentclass[prb,reprint]{revtex4-1} 

\usepackage{amsmath}
\usepackage{graphicx}

\newcommand{\dbar}{{d\mkern-6mu\mathchar'26}}
\newcommand{\up}{\uparrow}
\newcommand{\down}{\downarrow}

\begin{document}

\title{A different approach to introducing statistical mechanics}

\author{Thomas A. Moore}
\email{tmoore@pomona.edu}
\affiliation{Department of Physics and Astronomy, Pomona College, Claremont, California 91711}

\author{Daniel V. Schroeder}
\email{dschroeder@weber.edu}
\affiliation{Department of Physics, Weber State University, Ogden, Utah 84408-2508}

\begin{abstract}
The basic notions of statistical mechanics (microstates, multiplicities)
are quite simple, but understanding how the second law arises from
these ideas requires working with cumbersomely large numbers.  To avoid
getting bogged down in mathematics, one can compute multiplicities 
numerically for a simple model system such as an Einstein solid---a 
collection of identical quantum harmonic oscillators.
A computer spreadsheet program or comparable software 
can compute the required combinatoric functions for systems
containing a few hundred oscillators and units of energy.  When two
such systems can exchange energy, one immediately sees that some 
configurations are overwhelmingly more probable than others.
Graphs of entropy vs.\ energy for the two systems can be used to motivate
the theoretical definition of temperature, $T= (\partial S/\partial U)^{-1},$
thus bridging the gap between the classical and statistical approaches
to entropy.  Further spreadsheet exercises can be used to compute the
heat capacity of an Einstein solid, study the Boltzmann distribution,
and explore the properties of
a two-state paramagnetic system.  \textsl{\copyright 1997 American Association
of Physics Teachers.}  Published in Am.\ J.\ Phys.\ \textbf{65} (1), 26--36 (1997), \url{<http://dx.doi.org/10.1119/1.18490>}.\end{abstract}

\maketitle

\section{INTRODUCTION}

Entropy is a crucial concept in thermal physics.  A solid understanding
of what entropy is and  how it works is the key to understanding a broad
range of physical phenomena, and enhances  one's understanding of an even
wider range of phenomena. The idea of entropy and its (nearly)  inevitable
increase are thus core concepts that we should hope physics students at
all levels would  study carefully and learn to apply 
successfully.\cite{Andrew1984}

Unfortunately, the approach to teaching entropy found in virtually all
introductory and many  upper-level textbooks is more abstract
than it needs to be.  Such texts\cite{IntroTexts} 
introduce entropy in the context of macroscopic
thermodynamics, using heat engines and Carnot cycles to motivate the 
discussion, ultimately defining entropy to be the quantity that increases
by $Q/T$ when energy $Q$ is transferred by heating to a system at temperature~$T$.
In spite of the way that this approach mirrors the historical
development of the idea of entropy, our students have usually
found the intricate logic and abstraction involved in getting to this
definition too subtle to give them any intuitive
sense of what entropy is and what it measures; the concept ultimately
remains for them simply a formula to memorize.\cite{Clausius}
Most introductory texts  attempt to deal with this problem
by including a section about how entropy quantifies ``disorder''.  The
accuracy and utility of this idea, however, depends on
exactly how students  visualize ``disorder'', and in many situations this
idea is unhelpful or even actively misleading.\cite{disorder}  The
macroscopic approach  therefore generally does {\em not\/} help students develop
the kind of accurate and useful intuitive  understanding of this very
important concept that we would like them to have. 

The {\em statistical\/} interpretation of entropy (as the logarithm
of the the number of quantum  microstates consistent with a system's
macrostate) is, comparatively, quite concrete and  straightforward. The
problem here is that it is not trivial to show from this definition
that the entropy of a macroscopic system of interacting objects
will inevitably increase.\cite{Boltzmann}

This paper describes our attempt to  link the
statistical definition of entropy to the second law of thermodynamics in a
way that is as  accessible to students as possible.\cite{Baierlein}  Our primary motivation
was to create a more contemporary  approach to entropy to serve as the
heart of one of the six modules for {\it Six Ideas That Shaped  Physics}, a
new calculus-based introductory course developed with the support and
assistance of  the Introductory University Physics Project (IUPP).  However, we
have found this approach to be equally  useful in upper-level undergraduate
courses.

The key to our approach is the use of a {\em computer\/} to count
the numbers of  microstates associated with various macrostates of a
simple model system.  
While the processes  involved in the calculation are
simple and easy to understand, the computer makes it practical to do  the
calculation for systems large enough to display the irreversible behavior
of macroscopic  objects.  This vividly displays the link between the
statistical definition of entropy and the second  law without using
any sophisticated mathematics. 
Our approach has proved to be very helpful to upper-level
students and is the basis of one of the most  successful sections of the 
{\it Six Ideas\/}
introductory course.\cite{SixIdeas}
Depending on the level of the course and the time available, the
computer-based approach can be used either as the primary tool or as a
springboard to the more powerful analytic methods presented in statistical
mechanics texts.

A number of choices are available for the specific
computational environment:  a standard spreadsheet program such as
Microsoft {\it Excel\/} or Borland {\it QuattroPro\/}; a 
stand-alone program specially written for the purpose; or even a 
graphing calculator such as the TI-82.  We will use spreadsheets in
our examples in the main body of this paper, and briefly describe
the other alternatives in Section~VI.

After some preliminary definitions in Section II, we present the core
of our approach, the consideration of two interacting ``Einstein
solids'', in Section III.  In the remaining sections we derive the
relation between entropy, heat, and temperature, then go on to
describe a number of other applications of spreadsheet
calculations in statistical mechanics.

\section{PHYSICS BACKGROUND}

\subsection{Basic concepts of statistical physics}

The aim of statistical mechanics is to explain the 
behavior of  systems comprised of a very large number of
particles.  We can characterize
the physical {\it state\/} of such a system in either of two ways, depending
on  whether we focus on its characteristics at the macroscopic or
molecular level.

The {\it macrostate\/} of a system is its state as viewed at a macroscopic
level.  For instance, to describe the macrostate of a tank of helium gas,
we could specify the number of molecules $N$ it contains, the total volume
$V$ that they occupy, and their total internal energy~$U$.  As long as the
gas is in internal thermal equilibrium, these three parameters suffice to
determine its macrostate.  (Other thermodynamic variables, such as the
temperature $T$ and the pressure $P$ of the gas, can then be calculated using
various ``equations of state'' such as the ideal gas law $PV=NkT$ and the
relation $U={3\over2}NkT$, where $k$ is Boltzmann's constant.) In general, the
macrostate of a system is determined by the various conservation laws and
physical constraints that it is subject to.

The {\it microstate\/} of a system is its state as viewed at the molecular
level.  To describe the microstate of our tank of helium gas, we would
have to specify the exact position and velocity (or quantum mechanical
wavefunction) of each individual molecule.  Note that if we know a
system's microstate, we also know its macrostate:  The total energy $U$,
for instance, is just the sum of the energies of all the molecules.  On
the other hand, knowing a system's macrostate does not mean we're
anywhere close to knowing its microstate.

The number of microstates that are consistent with a given macrostate is
called the {\it multiplicity\/} of that macrostate, denoted $\Omega$. 
Equivalently, $\Omega$ is the number of {\it accessible\/} microstates, limited
only by the constraints that determine the macrostate.   For our tank of
helium gas, the multiplicity would be a function of three macroscopic
variables, for instance, $\Omega(U,V,N)$.  In most cases the multiplicity
will be a {\em very\/} large number, since there will be an enormous number of
different ways to distribute a given amount of energy among the system's
molecules (and, for fluids, to distribute the molecules within the
available volume).

The {\it fundamental assumption\/} of statistical mechanics is that an isolated
system in a given macrostate is equally likely to be in any of its
accessible microstates.  This means that the probability of finding the
system in any given microstate is $1/\Omega$, since there are $\Omega$
microstates, all equally probable.  In the next section we will see how
this assumption leads to the conclusion that macroscopic objects must
exhibit {\em irreversible\/} behavior, such as the spontaneous flow of energy
from a hot object to a cold one. In order to make this argument concrete,
though, we will focus our attention on a  specific model for a
thermodynamic system, since the exact way that a macrostate's
multiplicity  depends on $N$, $U$, and $V$ depends on what physical
system we are  considering.  Once we understand how the argument works
for our specific model, we will be able to understand (at least
qualitatively) how it would apply to other physical systems.

\subsection{The Einstein solid model}

There are only a few realistic systems whose multiplicities can
be calculated using elementary methods.  Even an ideal gas is too
complicated in this regard---a direct derivation of its multiplicity
function requires some fairly sophisticated tools.\cite{StoweNartonis}
A much simpler system
would be one whose particles are all fixed in place (so the
volume is irrelevant), storing energy
in units that are all the same size.  This idealized model is treated
in many statistical mechanics texts.\cite{ESolidRefs}
Physically, it corresponds to a collection of identical
quantum harmonic oscillators.  In 1907, Einstein proposed such a model
to describe the thermal properties of a simple crystalline solid,
treating each atom as an independent three-dimensional harmonic
oscillator.  We will therefore call this model an {\it Einstein
solid}.

We will use $N$ to denote the number of one-dimensional {\em oscillators\/} 
in an Einstein solid.  (So the number of atoms is $N/3$.)  If we neglect
the static zero-point energy (which plays no role in the thermal behavior),
then the energy of any oscillator can be written
\begin{equation}
E = \epsilon n,    \label{subtractedE}
\end{equation}
where $n$ is a nonnegative integer and $\epsilon$ is a constant
($\epsilon = hf$, where $f$ is the oscillator's natural
frequency).  In other words, each oscillator can store any number of
units of energy, all of the same size.  The total energy of the entire
solid can thus be written
\begin{equation}
U = \epsilon q,   \label{totalU}
\end{equation}
where $q$ is another nonnegative integer.  The multiplicity of an
Einstein solid is some function of $N$ and $q$:  the number of ways
of arranging $q$ units of energy among $N$ oscillators.

\begin{figure*}[t]
\centering
\includegraphics[width=12cm]{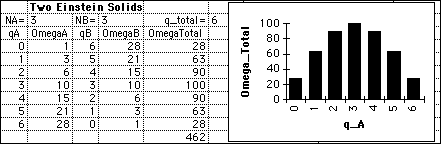}
\caption{Computation of the
multiplicies for two Einstein solids, each containing three oscillators,
which share a total of six units of energy.  The individual
multiplicities ($\Omega_A$ and $\Omega_B$) can be computed by simply counting
the various  microstates, or by applying Eq.~(\ref{EsolidMult}).  The total
multiplicity $\Omega_{\rm total}$ is the product of the individual
multiplicities, because there are $\Omega_B$ microstates of solid $B$ for
{\em each\/} of the $\Omega_A$ microstates of  solid $A$.  The total number of
microstates (462) is computed as the sum of the last column, but can be
checked by applying Eq.~(\ref{EsolidMult}) to the combined system of six
oscillators and six energy units.  The final column is plotted in the
graph at right.}
\end{figure*}

For a very small Einstein solid one can simply {\em count\/} all
the ways of arranging the energy among the oscillators.
For example, if $N = q = 3$, the
following 10 distinct distributions are possible (where the first number
in each triplet  specifies the number of energy units in the first
oscillator, the second the number of  energy units in the
second oscillator, etc.): 300, 030, 003, 210, 120, 201, 102, 021, 012,
111.  Thus the multiplicity of this macrostate is $\Omega = 10$.  By directly
counting microstates in this  manner, one can
not only vividly illustrate the concepts of macrostates  and microstates
to first-year students, but also show how the multiplicity increases
very  sharply with both $N$ and $q$, without using any difficult mathematics.

The general formula for the multiplicity of an Einstein solid is
\begin{align}
\Omega(N,q) = { q+N-1 \choose q } &= {(q+N-1)! \over q!(N-1)! }\nonumber \\
&=  {(U/\epsilon + N - 1)!  \over  (U/\epsilon)!(N - 1)!}.
\label{EsolidMult}
\end{align}
An elegant derivation of this formula can be found in the textbook
of Callen.\cite{CallenDerivation}  
In brief, the trick is to represent any microstate as
a sequence of dots (representing energy units) and lines (representing
partitions between one oscillator and the next).  A total of $q$ dots
and $N-1$ lines are required, for a total of $q+N-1$ symbols.  The
number of possible arrangements is just the number of ways of choosing
$q$ of these symbols to be dots.

Whether or not one derives formula (\ref{EsolidMult}) in class, students
have little trouble believing it after they have verified a few simple
cases by brute-force counting.  In an introductory course, this is
all we usually have them do.

\section{INTERACTING SYSTEMS}

To understand irreversible processes and the second law, we now move on
to consider a system of {\em two\/} Einstein solids, $A$ and $B$, which are free
to  exchange energy with each other but isolated from the rest of the
universe.

First we should be clear about what is meant by the ``macrostate'' of
such a composite system.  For simplicity, we assume that our two solids
are {\em weakly coupled}, so that the exchange of energy between them is
much slower than the exchange of energy among atoms within each solid. 
Then the individual  energies of the two solids, $U_A$ and $U_B$, will
change only slowly; over sufficiently short time scales they are
essentially fixed.  
We will use the word ``macrostate'' to refer to the
state of the combined system, as specified by the (temporarily)
constrained values of $U_A$ and $U_B$.  For any such macrostate we can
compute the multiplicity, as we shall soon see.  However, on longer time
scales the values of $U_A$ and $U_B$ {\em will\/} change, so we will
also talk  about the {\em total\/} multiplicity of the combined
system, considering all microstates to be ``accessible'' with only
$U_{\rm total} = U_A + U_B$ held fixed.  In any case, we must also specify the
number of oscillators in each solid, $N_A$ and $N_B$, in order to fully
describe the macrostate of the system.

Let us start with a very small system,
in which each of the ``solids'' contains only three oscillators
and the solids share a total of six units of energy:\cite{WhitneyBeatUs}
\begin{equation}
\begin{aligned}
  N_A &= N_B = 3; \\
  q_{\rm total} &= q_A + q_B = 6.
\end{aligned} \label{smallsystem}
\end{equation}
Given these parameters, we must still specify the individual
value of $q_A$ or $q_B$ to describe the macrostate of the system.
There are seven possible macrostates, with $q_A = 0$, 1, $\ldots$ 6.  
For each of these macrostates, we can compute the multiplicity of the 
combined system as the product of the multiplicities of the individual
systems:
\begin{equation}
  \Omega_{\rm total} = \Omega_A \Omega_B.  \label{multmult}
\end{equation}
This works because the systems are independent of each other:  For each
of the $\Omega_A$ microstates accessible to solid $A$, there 
are $\Omega_B$
microstates  accessible to solid $B$. A table of the various possible
values of $\Omega_A$, $\Omega_B$, and $\Omega_{\rm total}$ is shown in 
Fig.~1.  Students can
easily construct this table for themselves, computing the 
multiplicities by brute-force counting if necessary.  

\begin{figure*}[t]
\centering
\includegraphics[width=14cm]{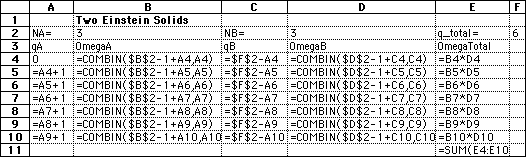}
\caption{Formulas from the {\it Excel\/} spreadsheet that produced
Fig.~1.  (Most of the formulas can be generated automatically by
copying from one row to the next.  Dollar signs indicate ``absolute'' references, which
remain unchanged when a formula is copied.)}
\end{figure*}

From the table we see that the fourth macrostate, in which each solid
has three  units of energy, has the highest multiplicity (100).  Now let
us invoke the  fundamental assumption:  On long time scales, {\em all\/} of
the 462 microstates are  accessible to the system, and therefore, by
assumption, are equally probable. The probability of finding the system
in any given {\em macro\/}state is therefore proportional to that macrostate's
multiplicity.  For instance, in this example the fourth macrostate is
the most probable, and we would expect to find the system in this
macrostate 100/462 of the time.  The macrostates in which one solid  or
the other has all six units of energy are less probable, but not 
overwhelmingly so.

Another way to construct the table of Fig.~1 is to use a computer
spreadsheet  program.  Both Microsoft {\it Excel\/} and Borland
{\it QuattroPro\/} include built-in functions for
computing binomial coefficients:
\begin{equation}
{q{+}N{-}1 \choose q}  = 
\begin{cases}
 {\tt COMBIN(q{+}N{-}1,q)}  & \text{in {\it Excel\/}}; \\
        {\tt MULTINOMIAL(q,N{-}1)} & \text{in {\it QuattroPro}}.
\end{cases}
\label{SSfunctions}
\end{equation}
The formulas from the {\it Excel\/} spreadsheet that produced Fig.~1 are shown
in Fig.~2.

The advantage in using a computer is that it is now easy to change the
parameters to consider larger systems.  Changing the numbers of
oscillators in the two solids can be accomplished by merely typing new
numbers in cells B2 and D2.  Changing the number of energy units
requires also adding more lines to the spreadsheet.  The spreadsheet
becomes rather cumbersome with much more than a few hundred energy units,
while overflow errors can occur if there are more than a few thousand
oscillators in each solid.  Fortunately, most of the important
qualitative features of thermally interacting systems can be observed
long before these limits are reached.

\begin{figure*}[t]
\centering
\includegraphics[width=15cm]{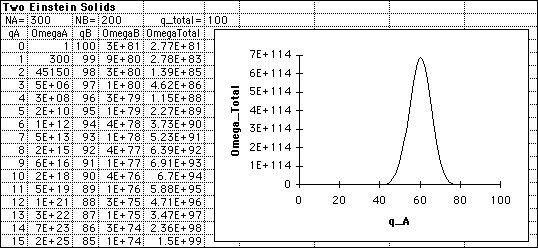}
\caption{Spreadsheet calculation of the multiplicity of
each macrostate, for two larger Einstein solids sharing 100 energy
units.  Here solids $A$ and $B$ contain 300 and 200 energy units,
respectively.  Only the first few lines of the table are shown; the
table continues down to $q_A=100$.  The final column is plotted in the 
graph at right.  Notice from the graph that the multiplicity at $q_A=60$ is
more than $10^{33}$ times larger than at $q_A=0$.}
\end{figure*}

Figure 3 shows a spreadsheet-generated table, and a plot of $\Omega_{\rm total}$
vs.\ $q_A$, for the parameters
\begin{equation}
N_A = 300,\quad  N_B = 200,\quad  q_{\rm total} = 100.\label{largersolids}
\end{equation}
We see immediately that the multiplicities are now quite large.  But more
importantly, the multiplicities of the various macrostates now differ by
{\em factors\/} that can be very large---more than $10^{33}$ in the most
extreme case.  If we again make the fundamental assumption, this result
implies that the most likely macrostates are $10^{33}$ times more probable
than the least likely macrostates.  If this system is initially in one
of the less probable macrostates, and we then put the two solids in
thermal contact and wait a while, we are almost {\em certain\/} to find it
fairly near the most probable macrostate of $q_A=60$ when we check again
later.  In other words, this system will exhibit irreversible behavior,
with net energy flowing spontaneously from one solid to the other, until
it is near the most likely macrostate.  This conclusion is
just a statement of the second law of thermodynamics.

There is one important limitation in working with systems of only a few
hundred particles.  According to the graph of Fig.~3, there are actually
quite a few macrostates with multiplicities nearly as large as that of the most
likely macrostate. In relative terms, however, the width of this peak
will decrease as the  numbers of oscillators and energy units
increase.  This is already apparent in comparing Figs.\ 1 and~3.  If the
numbers were increased further, we would eventually find that the
relative width of the peak (compared to the entire horizontal scale of
the graph) is completely negligible.  Unfortunately, one cannot verify
this statement directly with the spreadsheets.  Students in an
introductory class have little trouble believing this result, when
reminded of the fact that macroscopic systems contain on order of
$10^{23}$ particles and energy units.  In an upper-division course, one
can prove using Stirling's approximation that the relative width of the
peak is of order $1/\sqrt N$ or $1/\sqrt q$, whichever is larger.  Thus,
statistical fluctuations away from the most likely macrostate will be
only about one part in $10^{11}$ for macroscopic systems.\cite{Baierlein1978}

Using this result, we can state the second law more strongly:  Aside
from  fluctuations that are normally much too small to measure, any isolated
macroscopic system will inevitably evolve toward its {\em most\/} likely
accessible macrostate.

Although we have proved this result only for a system of two weakly
coupled Einstein solids, similar results apply to all other macroscopic
systems.  The details of the counting become more complex when the
energy units can come in different sizes, but the essential properties
of the combinatorics of large numbers do not depend on these details.
If the subsystems are not weakly coupled, the division of the
whole system into subsystems becomes a mere mental exercise with
no real physical basis; nevertheless, the same counting arguments
apply, and the system will still exhibit irreversible behavior if
it is initially in a macrostate that is in some way ``inhomogeneous''.

\section{ENTROPY AND TEMPERATURE}

Instead of always working with multiplicities, it is convenient to
define the {\it entropy\/} of a system as
\begin{equation}
  S = k \ln \Omega.   \label{Sdef}
\end{equation}
The factor of Boltzmann's constant is purely conventional, but the
logarithm is important:  It means that, for any given macrostate,
the total entropy of two weakly interacting\cite{TimeScaleNote}
systems is the {\em sum\/} of their individual entropies:
\begin{align}
  S_{\rm total} &= k \ln(\Omega_A \Omega_B)\nonumber\\
          &= k \ln \Omega_A + k \ln \Omega_B\nonumber\\
          &= S_A + S_B.     \label{Sadditivity}
\end{align}
Since entropy is a strictly increasing function of multiplicity,
we can restate the second law as follows:  Aside from 
fluctuations that are normally much too small to measure, any isolated macroscopic
system will inevitably evolve toward whatever (accessible) macrostate
has the largest entropy.

It is a simple matter to insert columns for $S_A$, $S_B$, and $S_{\rm total}$ into
the spreadsheet in Fig.~3.  A graph of these three quantities
is shown in Fig.~4.  Notice that $S_{\rm total}$ is {\em not\/} a strongly peaked
function.  Nevertheless, the analysis of the previous section tells
us that, when this system is in equilibrium, it will almost certainly
be found in a macrostate near $q_A = 60$.

\begin{figure}[b]
\centering
\includegraphics[width=8cm]{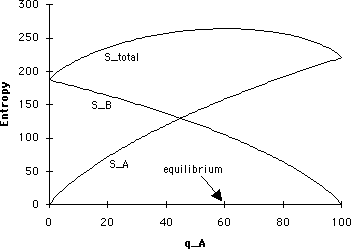}
\caption{A spreadsheet-generated plot of $S_A$, $S_B$, and $S_{\rm total}$ for
the system of two Einstein solids represented in Fig.~3.  At equilibrium,
the slopes of the individual entropy graphs are equal in magnitude.
Away from equilibrium, the solid whose slope $dS/dq$ is {\em steeper\/} tends
to gain energy, while the solid whose slope $dS/dq$ is {\em shallower\/} tends
to lose energy.}
\end{figure}

Now let us ask how entropy is related to temperature.  The most fundamental
way to define temperature is in terms of energy flow and thermal equilibrium:
Two objects in thermal contact are said to be at the same temperature if 
they are in thermal equilibrium, that is, if there is no spontaneous
net flow of energy between them.  If energy {\em does\/} flow spontaneously
from one to the other, then we say the one that loses energy has the
higher temperature while the one that gains energy has the lower temperature.

Referring again to Fig.~4, it is not hard to identify a quantity
that is the same for solids $A$ and $B$ when they are in equilibrium.
Equilibrium occurs when $S_{\rm total}$ reaches a maximum, that is, when the
slope $dS_\textrm{total}/dq_A = 0$.  But this slope is the sum of the slopes of
the individual entropy graphs, so their slopes must be equal in magnitude:
\begin{equation}
  {dS_A\over dq_A} = - {dS_B\over dq_A}\qquad\hbox{at equilbrium}. 
\label{SlopesEqual}
\end{equation}
If we were to plot $S_B$ vs.~$q_B$ instead of $q_A$, its slope at the equilbrium
point would be equal, in sign and magnitude, to that of $S_A(q_A)$.
So the quantity that is the same for both solids when they are in
equilibrium is the slope $dS/dq$.  Temperature must be some function
of this quantity.\cite{EntropiesAdd}

To identify the precise relation between temperature and the slope
of the entropy vs.\ energy graph, look now to a point in Fig.~4 where
$q_A$ is larger than its equilibrium value.  Here the entropy graph
for solid $B$ is steeper than that for solid $A$, meaning that if a bit
of energy were to pass from $A$ to $B$, solid $B$ would gain more entropy
than solid $A$ loses.  Since the {\em total\/} entropy would increase, the
second law tells us that this process will happen spontaneously.
In general, the steeper an object's entropy vs.\ energy graph, the
more it ``wants'' to gain energy (in order to obey the second law), 
while the shallower an object's entropy vs. energy graph, the less
it ``minds'' losing a bit of energy.  We therefore conclude that
temperature is {\em inversely\/} related to the slope $dS/dU$.  In fact,
the reciprocal $(dS/dU)^{-1}$ has precisely the units of temperature,
so we might guess simply
\begin{equation}
  {1\over T} = {dS\over dU}.     \label{TempDef}
\end{equation}
The factor of Boltzmann's constant in the definition (\ref{Sdef}) of entropy
eliminates the need for any further constants in this formula.  To confirm
that this relation gives temperature in ordinary Kelvin units one
must check a particular example, as we will do in the following section.

Our logical development of the concepts of entropy and temperature from
statistical considerations is now complete.  In the following section,
we will apply these ideas to the calculation of heat capacities and
other thermal properties of some simple systems.  However, at this
point in the development, many 
instructors will wish to first generalize the discussion to 
include systems (such as gases) whose volume can change.  The derivatives in 
Eqs.\ (\ref{SlopesEqual}) and (\ref{TempDef}) then become partial derivatives at fixed volume, and
one cannot simply rearrange Eq.~(\ref{TempDef}) to obtain $dS = dU/T$.  There are at
least two ways to get to the correct relation between entropy and
temperature, and we now outline these briefly.

The standard approach in statistical mechanics texts\cite{Mandl}
is to first consider two interacting systems separated
by a movable partition, invoking the second law to identify pressure by the 
relation $P = T(\partial S/\partial V)_U$.  Combining this 
relation with Eq.~(\ref{TempDef}) one obtains 
the thermodynamic identity $dU = TdS + (-PdV)$.  For quasistatic processes
(in which the change in volume is sufficiently gradual), the
second term is the (infinitesimal) work $W$ done on the system.\cite{InexactDifferentials}
But the first law says that for any process, 
$dU = Q + W$, where $Q$ is the (infinitesimal) energy added by heating.  Therefore, for 
quasistatic processes only, we can identify $TdS = Q$ or
\begin{equation}
   dS = {Q\over T}\qquad \hbox{(quasistatic processes only)}.
\label{SQrelation}
\end{equation}
From this relation one can go on to treat a variety of applications,
most notably engines and refrigerators.

In an introductory course, however, there may not be time for the
general discussion outlined in the previous paragraph.  Fortunately,
if one does not need the thermodynamic identity, some of the steps
can be bypassed (at some cost in rigor).  We have already seen that,
when the volume of a system does not change, its entropy changes by
\begin{equation}
   dS = {dU\over T}\qquad \hbox{(constant volume)}. \label{dSatConstV}
\end{equation}
Now consider a system whose volume does change, but that is otherwise
isolated from its environment.  If the change in volume is quasistatic
(i.e., gradual), then the entropy of the system cannot change.  Why?  
Theoretically, because while the energies of the system's quantum states 
(and of the particles in those states) will be shifted, the 
process is not violent enough to knock a particle from one quantum state
to another; thus the multiplicity of the system remains fixed.  Alternatively,
we know from experiment that by reversing the direction of an adiabatic,
quasistatic volume
change we can return the system to a state arbitrarily close to its initial state; 
if its entropy 
had increased, this would not be possible even in the quasistatic limit.  
By either argument, the work
done on a gas as it is compressed should not be included in
the change in energy $dU$ that contributes to the change in entropy.
The natural generalization of Eq.~(\ref{dSatConstV}) to non-constant-volume
quasistatic processes is therefore to replace $dU$ with $dU - W = Q$, the
energy added to the system by heating {\em alone}.  
Thus we again arrive at Eq.~(\ref{SQrelation}).

\section{FURTHER SPREADSHEET CALCULATIONS}

So far in this paper we have used spreadsheet calculations only
to gain a qualitative understanding of the second law and the
relation between entropy and temperature.  Now that we have
developed these concepts, however, we can use the same tools
to make quantitative predictions about the thermal behavior of various systems.

\begin{figure*}
\centering
\includegraphics[width=14cm]{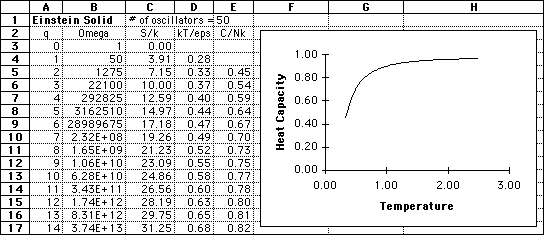}
\caption{Spreadsheet calculation of the entropy, temperature, and heat capacity
of an Einstein solid.  (The table continues down to $q=100$.)
The graph at right shows the heat capacity as a
function of temperature.  To see the behavior at lower temperatures
one can simply increase the number of oscillators in cell~E1.}
\end{figure*}

\subsection{Heat capacity of an Einstein solid}

Figure 5 shows a spreadsheet calculation of the entropy of a 
{\em single\/} Einstein solid of $N=50$ oscillators with anywhere from 0 to 100
units of energy.  Then, in the fourth column, we have calculated the
temperature  $T = \Delta U / \Delta S$  using a pair of closely spaced
entropy and energy values.  We have actually used a centered-difference
approximation, so the cell D4, for example, contains the formula
$\tt 2 / (C5 - C3)$.  Since we have omitted the constants $k$ and $\epsilon$ from
the formulas, the temperatures are expressed in units of $\epsilon/k$.

We now know the temperature for every energy, or vice versa.  In
practice, however, one usually compares such predictions to experimental
measurements of the {\it heat capacity}, defined (for systems at constant
volume) as 
\begin{equation}
  C = {dU\over dT}\qquad   \hbox{(constant volume)}.  \label{Cdef}
\end{equation}
This quantity, normalized to the number of oscillators, is computed in the 
fifth column of the table, again using a centered-difference approximation 
(for instance, cell E5 contains the formula $\tt(2/(D6-D4))/\$E\$1$).  Since we
are still neglecting dimensionful constants, the entries in the column are
actually for the quantity $C/(Nk)$.

To the right of the table is a spreadsheet-generated graph of the heat
capacity vs.\ temperature for this system.  This theoretical prediction
can now be compared to experimental data, or to the graphs often given
in textbooks.\cite{HRWfigure}  The quantitative agreement with experiment
verifies that
the formula $1/T=dS/dU$ is probably correct.\cite{TempDefNote}
At high temperatures we find
the constant value $C = Nk$, as expected from the equipartition theorem
(with each oscillator contributing two degrees of freedom).  At temperatures
below $\epsilon/k$, however, the heat capacity drops off sharply.  This is
just the qualitative behavior that Einstein was trying to explain with
his model in 1907.  

To investigate the behavior of this system at lower temperatures, one can
simply increase the number of oscillators in cell E1.  With 45000 oscillators 
the temperature goes down to about $0.1\epsilon/k$, corresponding to a heat 
capacity per oscillator of about $.005 k$.  At this point it is apparent
that the curve becomes {\em extremely\/} flat at very low temperatures.  
Experimental measurements on real solids confirm that the heat capacity
goes to zero at low temperature, but do not confirm the detailed shape
of this curve.  A more accurate treatment
of the heat capacities of solids requires more sophisticated tools, but
is given in standard textbooks.\cite{DebyeTheory}

A collection of identical harmonic oscillators is actually a more accurate
model of a completely different system:  the vibrational degrees of freedom
of an ideal diatomic gas.  Here again, experiments confirm that the
equipartition theorem is satisfied at high temperatures but the vibration
(and its associated heat capacity) ``freezes out'' when $kT \ll \epsilon$.
Graphs showing this behavior are included in many textbooks.\cite{HRWgasFig}

\subsection{The Boltzmann distribution}

Perhaps the most useful tool of statistical mechanics is the
{\it Boltzmann factor}, $e^{-E/kT}$, which gives the relative probability
of a system being in a microstate with energy~$E$, when it is in
thermal contact with a much larger system (or ``reservoir'')
at temperature~$T$.  A fairly straightforward derivation of this result 
can be found in most textbooks on statistical mechanics.\cite{BoltzmannFactorDerivations}  
A less general derivation,
based on the exponentially decreasing density of an isothermal atmosphere,
is given in Serway's introductory text.\cite{SerwayBoltzmann}

\begin{figure*}
\centering
\includegraphics[width=14cm]{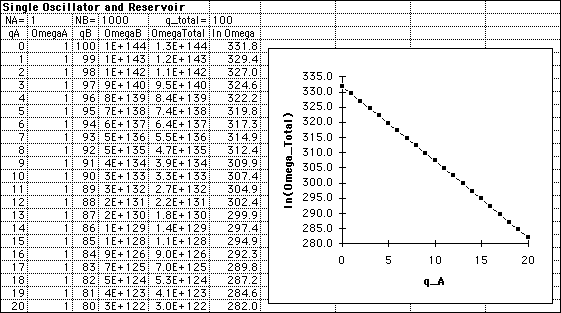}
\caption{States of a single oscillator in thermal contact with a large reservoir
of identical oscillators.  A graph of the logarithm of the multiplicity
vs.\ the oscillator's energy has a constant, negative slope, in accord
with the Boltzmann distribution.}
\end{figure*}

Although it is no substitute for an analytic treatment, a spreadsheet
calculation can easily display
this exponential dependence for the case of a 
single one-dimensional oscillator in thermal contact with a large Einstein 
solid.\cite{WhitneyBoltzmann}
Figure~6 shows such a calculation, for the parameters $N_A = 1$, 
$N_B = 1000$, and $q_{\rm total}=100$.  Here solid $A$ is the ``system'',
while solid $B$, which is much larger, is the ``reservoir''. 
The graph shows the natural logarithm of the total multiplicity, plotted 
as a function of $q_A$.  According to the fundamental assumption, the
total multiplicity is proportional to the probability of finding the
small system in each given microstate.  Note that the graph is a straight 
line with a negative slope:  This means that the probability
does indeed decay exponentially (at least approximately) as the energy 
$E = \epsilon q_A$ of the small system increases.

According to the Boltzmann factor, the slope of the graph of multiplicity
(or probability) vs.\ $q_A=E/\epsilon$ should equal $-\epsilon/kT$.
Since the small system always has unit multiplicity and zero entropy, 
the entropy of the combined system is the same as the entropy 
of the reservoir: $S_{\rm total} = S_B$.  Once they know this, students 
can check either numerically or analytically that the slope of the graph 
is 
\begin{equation}
{\partial\over\partial q_A}\ln\Omega_{\rm total} 
= {\epsilon\over k} {\partial S_B \over \partial U_A} 
= - {\epsilon\over k} {\partial S_B\over\partial U_B}
= -{\epsilon\over k T_B},
\end{equation}
where the derivatives and the temperature $T_B$ are all evaluated 
at $U_A = 0$.

\subsection{Thermal behavior of a two-state paramagnet}

We have used the Einstein solid model in our examples for several reasons:
it is fairly realistic, its behavior is intuitive, and, most importantly,
it is mathematically simple.  Other systems generally lack at least one
of these qualities.  To carry out our spreadsheet calculations, however,
all we require is mathematical simplicity.

The simplest possible system would consist of ``atoms'' that have only 
two allowed energy levels.  A common example is a paramagnetic material,
with one unpaired electron per atom, immersed in a uniform magnetic field $B$.
Let us say that $B$ points in the $+z$ direction, and quantize the spin
(hence the magnetic moment) of the atom along this axis.  When the
magnetic moment points up, the atom is in the ground state, while when
the magnetic moment points down, the atom is in the excited state.
The energy levels of these two states are
\begin{equation}
\begin{aligned}
    {\rm up}&: \ E_\up   = - \mu B,\\
  {\rm down}&: \ E_\down = + \mu B,
\end{aligned}  \label{magnetEs}
\end{equation}
where $\mu$ is the magnitude of the atom's magnetic moment.  In a collection 
of $N$ such atoms, the total energy will be
\begin{equation}
U = N_\up E_\up + N_\down E_\down = \mu B (N_\down - N_\up),  
\label{magnetTotalE}
\end{equation}
where $N_\up$ and $N_\down$ are the numbers of atoms in the up and down states.
These numbers must add up to $N$:
\begin{equation}
N_\up + N_\down = N.  \label{totalN}
\end{equation}
The total magnetization of the sample is given by
\begin{equation}
M = \mu N_\up - \mu N_\down = \mu(N_\up - N_\down) = -E/B.
\end{equation}

\begin{figure*}
\centering
\includegraphics[width=14cm]{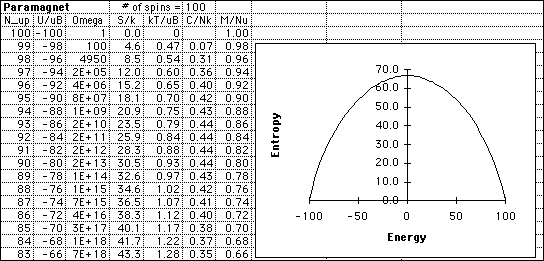}
\caption{Spreadsheet calculation of the entropy,
temperature, heat capacity, and magnetization of a two-state paramagnet.
(The symbol {\tt u} represents $\mu$, the magnetic moment.)
At right is a graph of entropy vs.\ energy, showing that as energy is
added, the temperature (given by the reciprocal of the slope) goes to infinity
and then becomes negative.}
\end{figure*}

To describe a macrostate of this system we would have to specify the
values of $N$ and $U$, or equivalently, $N$ and $N_\up$ (or $N$ and $N_\down$).
The multiplicity of any such macrostate is the number of ways of
choosing $N_\up$ atoms from a collection of $N$:
\begin{equation}
\Omega(N,N_\up) = {N \choose N_\up} = {N! \over N_\up! (N-N_\up)!}
                 = {N! \over N_\up! N_\down!}.  \label{magnetMult}
\end{equation}
From this simple formula we can calculate the entropy, temperature,
and heat capacity.

Figure 7 shows a spreadsheet calculation of these quantities for
a two-state paramagnet of 100 ``atoms''.  The rows of the table are
given in order of increasing energy, which means decreasing $N_\up$
and decreasing magnetization.  Alongside the table is a graph of
entropy vs.\ energy.  Since the multiplicity of this system is largest
when $N_\up$ and $N_\down$ are equal, the entropy also reaches a maximum at
this point ($U=0$) and then falls off as more energy is added.  This
behavior is unusual and counter-intuitive, but is a straightforward
consequence of the fact that each ``atom'' can store only a limited amount
of energy.

To appreciate the consequences of this behavior, imagine that the
system starts out with $U < 0$ and that we gradually add more energy.
At first it behaves like an Einstein solid:  The slope of its
entropy vs.\ energy graph decreases, so its tendency to suck in energy
(in order to maximize its entropy in accord with the second law)
decreases.  In other words, the system becomes hotter.  As $U$ approaches
zero, the system's desire to acquire more energy disappears.  Its
temperature is now infinite, meaning that it will willingly give up a
bit of energy to any other system whose temperature is finite (since
it loses no entropy in the process).  When $U$ increases further and becomes
positive, our system actually {\em wants\/} to give up energy.  Intuitively,
we would say that its temperature is higher than infinity.  But this
is mathematical nonsense, so it is better to think about the reciprocal
of the temperature, $1/T$, which is now lower than zero, in other words,
negative.  In these terms, our intuition even agrees with 
the precise relation (\ref{TempDef}).

The paramagnetic system can be an invaluable pedagogical tool, since
it forces students to abandon the naive notion that temperature is
merely a measure of the average energy in a system, and to embrace instead
the more general relation (\ref{TempDef}), with all its connotations.  
One can even imagine using the paramagnet in place of the Einstein
solid as the main paradigm for the lessons of Sections III and~IV.\cite{KKtwoState}  
In particular, it is not hard to set up a spreadsheet model of two interacting
paramagnets, to look at how energy can be shared between them.
However, this system exhibits behavior not commonly seen in other systems
or in students' prior experience.  We therefore prefer not to use it as
a first example, but rather to save it for an upper-level course in which
students have plenty of time to retrain their intuition.

\begin{figure*}
\centering
\includegraphics[width=12cm]{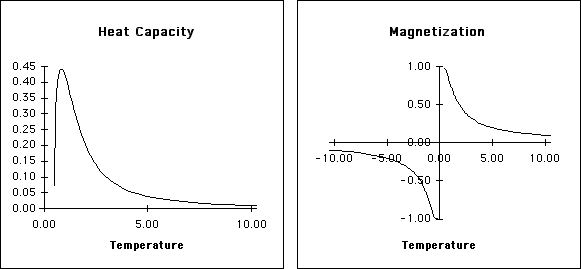}
\caption{Graphs of the heat capacity and magnetization as a function of
temperature for a two-state paramagnet, generated from the last three
columns of the spreadsheet shown in Fig.~7.}
\end{figure*}

Putting these fundamental issues aside, we can go on to calculate both the
heat capacity and the magnetization of our two-state paramagnet as
a function of temperature.  Plots of these quantities are shown in Fig.~8.
Since the heat capacity goes to zero at high temperature, the equipartition
theorem fails utterly to describe this system (as it must, since it
applies only to degrees of freedom for which the energy is a quadratic
function of a coordinate or momentum).  The magnetization is complete
at $T$ just greater than zero, but drops off as the temperature is
increased.  The behavior in the high-temperature limit is known as
Curie's law, as described in many textbooks.\cite{CuriesLaw}

\section{IMPLEMENTATION DETAILS}

The ideas presented in this paper have been tested, in various forms,
in calculus-based introductory courses at Pomona College, Grinnell
College, Smith College, Amherst College, and St.\ Lawrence University, and in
an upper-division thermodynamics course at Weber State University.
We have found that students at either level are quite able to understand
the fundamental concepts of statistical mechanics that we are trying
to teach.  In particular, the fact that the statistical approach to 
the second law is more ``modern'' does not, in our opinion, make it
unsuitable for first-year students.  On the contrary, we have had far
more success in teaching the statistical approach than in teaching
the concept of entropy from the older classical viewpoint used in most
texts.

The use of computers in our approach does not seem to pose a barrier
to today's students.  Many of them have had previous experience with
spreadsheet programs, and the rest are able to learn very quickly
with the help of a one-page instruction sheet that we have
prepared.\cite{InstructionSheet}

\subsection{Sample exercises}

We generally present all of the material of Section III in class, and
hand out reproductions of the figures from that section.
For homework, we ask students to reproduce Fig.~1, then change the parameters
to look at some larger systems.  Here is a typical homework assignment,
written for students who will be using {\sl Excel\/}:

\begin{enumerate}

\item List all the possible microstates for an Einstein solid
containing three oscillators and five units of energy.  Verify that
$\Omega = {5+3-1\choose5}$.

\item  Set up an {\sl Excel\/} spreadsheet identical (initially)
to the one shown on the handouts:  two Einstein solids, each
containing three harmonic oscillators, with a total of six units of energy.
You'll need to type in all the labels and formulas down to row~4,
as well as cell~A5.  After that, you can use the ``Fill Down''
command to fill in the rest (except for cell~E11).  Use the ``chart
wizard tool'' (the second icon from the right) to create the
graph.  Check that your spreadsheet yields all the same results
as on the handout.  Then modify your spreadsheet to show
the case where one Einstein solid contains six harmonic oscillators 
and the other contains four harmonic oscillators (with the total
number of energy units still equal to six).  Turn in a printout
of the modified spreadsheet.  Assuming that all microstates
are equally likely, what is the most probable macrostate,
and what is its probability?  What is the least probable macrostate,
and what is its probability?

\item  Modify your spreadsheet from the previous problem to show
the case where solid $A$ contains 100 oscillators, solid $B$
contains 200 oscillators, and there are 100 units of energy
in total.  It's easiest to delete the old graph and make a new
one, and this time it's best to make it an ``XY (scatter)'' graph
instead of a column graph.  Turn in a printout of the first several
rows of the spreadsheet as well as the graph.  What is the most probable
macrostate, and what is its probability?  What is the least probable
macrostate, and what is its probability?  Calculate the entropies
of these two macrostates (expressed as unitless numbers, neglecting
the factor of Boltzmann's constant).

\item  Starting with your spreadsheet from the previous problem,
add columns for the entropy of solid $A$, the entropy of solid $B$, and
the total entropy.  Print a graph showing all three entropies, plotted
vs.\ the energy of solid~$A$.  Use the formula $T=\Delta U/\Delta S$ to
calculate the temperatures of both solids (in units of $\epsilon/k$)
at the equilbrium point and also at the point $q_A = 60$.  Do your
results make sense?  Explain.

\end{enumerate}

We have also given computer-based homework assignments on the 
material of Section~V.  In upper-division courses we have 
asked students to generate Fig.~5 before seeing it in class.
The treatment of this material can vary greatly depending
on how the class is taught, however, so we will not reproduce our 
homework exercises here.

\subsection{Alternative computational environments}

Spreadsheet programs provide an almost ideal computational
environment for the calculations described in this paper:  they are versatile,
easy to learn, and readily available on most campuses.  Other 
computing options may be more suitable in some contexts, however.

Before we discovered spreadsheets, one of us (T.A.M.) wrote a stand-alone
program for the Macintosh to produce tables and graphs such as those
shown in Figs.\ 1 and~3.  The stand-alone program is even easier to use
than a spreadsheet, and can conveniently handle somewhat larger systems
and values of~$q$.  It also
allows one to completely omit formula (\ref{EsolidMult}) 
from the course if desired---students can simply be asked to verify the
computer's computations in a few simple cases.  This program is sufficient
to teach the core of our approach, as outlined in Section~III.  A copy
of the program is available on request.\cite{TomsProgram}

A much cheaper, but less convenient, option is to perform the calculations
on a modern graphing calculator.  We do not particularly recommend this; 
in our experience, graphing calculators are {\em much\/} harder to use than
spreadsheet programs.  However, a motivated student who already knows
how to use the calculator should be able to reproduce all the major results
in this paper without too much difficulty.  As an example, we
describe here an appropriate set of instructions for the TI-82.\cite{TI82overflow}

To set up tables like those shown in Section III, one can use the ``table''
feature of the TI-82, which displays seven rows at a time, calculating
each row as it is displayed.  The first column of the table, labeled {\tt X},
should be used for $q_A$.  The following formulas in the ``Y='' screen
will then generate the remaining four columns:

\begin{quote}
$\tt Y_{\rm 1} = (X+A-1)\ nCr\ X$\\
$\tt Y_{\rm 2} = Q-X$\\
$\tt Y_{\rm 3} = (Y_{\rm 2}+B-1)\ nCr\ Y_{\rm 2}$\\
$\tt Y_{\rm 4} = Y_{\rm 1} * Y_{\rm 3}$
\end{quote}

\noindent
Here we are using variables {\tt A}, {\tt B}, and {\tt Q} to hold
the constants $N_A$, $N_B$, and $q_{\rm total}$, respectively.
Before plotting a graph, it is best to set {\tt Xmin} to zero and
{\tt Xmax} to 94; this insures that each pixel corresponds to an
integer value of {\tt X}.

To calculate temperatures and heat capacities on the TI-82, one must
use ``lists'' instead of ``tables''.  The following sequence of instructions
will reproduce most of Fig.~5, storing $q$, $\Omega$, $S/k$, $kT/\epsilon$, 
and $C/Nk$ in lists {\tt L}$_1$ through {\tt L}$_5$:

\begin{quote}
$\tt seq(X,X,0,50,1)\rightarrow L_{\rm 1}$\\
$\tt seq((X+A-1)\ nCr X,X,0,50,1)\rightarrow L_{\rm 2}$\\
$\tt ln(L_{\rm 2})\rightarrow L_{\rm 3}$\\
$\tt seq(2/(L_{\rm 3}(X+2)-L_{\rm 3}(X)),X,1,49,1)\rightarrow L_{\rm 4}$\\
$\tt seq(2/(L_{\rm 4}(X+2)-L_{\rm 4}(X))/A,X,1,47,1)\rightarrow L_{\rm 5}$
\end{quote}

\noindent
Again we assume that the number of oscillators is stored in variable {\tt A}.
Lists 4 and 5 will be offset upward by one and two places respectively,
but this can be fixed by entering ``stat-edit'' mode and inserting zeros.  One
should also add zeros at the end to make all lists the same length.  It is
then a simple matter to plot the heat capacity vs.\ the temperature, as
shown in Fig.~5.  Unfortunately, the lists will not update automatically
when the value of {\tt A} is changed.


\begin{thebibliography}{99}
\bibitem{EntropiesAdd} Notice that the preceding argument depends crucially
on the additivity of $S=k\ln\Omega$ for the composite system.  Thus the
logarithm in the definition of $S$ is essential, if $S$ and $T$ are to be
related in a simple way.
\bibitem{TempDefNote} Most textbooks define
the temperature scale operationally in terms of a gas thermometer.  The most direct
verification of $1/T=dS/dU$ would therefore be to apply this formula directly to
an ideal gas.  But the gas thermometer can be used to measure the heat
capacity of a solid, and this result, in turn, can be compared to our
theoretical prediction.  Note, in particular, that the factor of Boltzmann's
constant in the definition of $S$ conveniently makes the temperature come
out in ordinary Kelvin units, with no further constants needed.
\bibitem{WhitneyBoltzmann} Similar calculations can be done by hand, albeit for
a somewhat smaller ``large'' system.  See Whitney, Ref.~\onlinecite{ESolidRefs}, p.\ 128.
\bibitem{TI82overflow} Since
the TI-82 overflows at $10^{100}$, one is limited to somewhat smaller systems
than we have used in our earlier examples.  In practice this is not much
of a limitation.
\end{thebibliography}
\end{document}